\documentclass[showpacs,twocolumn,prl]{revtex4}
\usepackage{amsmath,graphicx}
\usepackage{color,ulem}

\definecolor{darkred}{rgb}{0.90,0,0}
\definecolor{darkgreen}{rgb}{0,0.60,.2}
\definecolor{darkblue}{rgb}{0,0,1}
\definecolor{grey}{cmyk}{0,0,0,0.25}
\definecolor{orange}{cmyk}{0,0.6,0.8,0}

\begin{document}

\title{Conductance fluctuations and field asymmetry of rectification in graphene}
\author{C. Ojeda-Aristizabal}
\affiliation{LPS, Univ. Paris-Sud, CNRS, UMR 8502, F-91405 Orsay Cedex, France}
\author{M. Monteverde}
\affiliation{LPS, Univ. Paris-Sud, CNRS, UMR 8502, F-91405 Orsay Cedex, France}
\author{R.Weil}
\affiliation{LPS, Univ. Paris-Sud, CNRS, UMR 8502, F-91405 Orsay Cedex, France}
\author{M. Ferrier}
\affiliation{LPS, Univ. Paris-Sud, CNRS, UMR 8502, F-91405 Orsay Cedex, France}
\author{S. Gu\'eron}
\affiliation{LPS, Univ. Paris-Sud, CNRS, UMR 8502, F-91405 Orsay Cedex, France}
\author{H. Bouchiat}
\affiliation{LPS, Univ. Paris-Sud, CNRS, UMR 8502, F-91405 Orsay Cedex, France}

\begin{abstract}
We investigate conductance fluctuations as a function of carrier density $n$ and magnetic field in diffusive mesoscopic samples made from monolayer and bilayer graphene. We show that the fluctuations' correlation energy and field, which are functions of the diffusion coefficient, have fundamentally different variations with  $n$, illustrating the contrast between massive and massless carriers. 
The field dependent fluctuations are nearly independent of $n$, but the $n$-dependent fluctuations are not universal and are largest at the charge neutrality point. We also measure the second order conductance fluctuations (mesoscopic rectification). Its field asymmetry, due to electron-electron interaction, decays with conductance, as predicted for diffusive systems. 
\end{abstract}
\maketitle
Reproducible conductance fluctuations (CF) are one of the most striking signature of   phase coherent transport \cite{intro}. The  conductance of a mesoscopic sample results from interference between all wave packets  traversing the sample. This interference pattern  is sensitive to variations in disorder configuration, Fermi energy  or magnetic flux,  leading to reproducible CF as one of these parameters is changed. In diffusive or chaotic systems the CF amplitude has been shown  to be universal \cite{intro,lee87,web86} and ergodic,  i.e. independent of the mechanism of phase randomization (magnetic field, Fermi energy, configuration of impurities  for diffusive systems, sample shape for ballistic systems). The CF amplitude is of  the order of $e^2/h$, with a coefficient which only depends on the symmetry class of the mesoscopic system. The typical correlation energies $E_{\varphi}$ and fields $B_\varphi$ of the fluctuations depend upon the typical time $\tau_{int}$ and area $A_{int}$ over which interference occur: $E_{\varphi}=\hbar/\tau_{int}$ and $B_\varphi=\Phi_0/A_{int}$, with $\Phi_0=h/e$ \cite{lee87}.
CF have been extensively investigated in metallic and semiconducting systems \cite{intro,web86}. The recently discovered graphene \cite{geim} provides a unique system in which the Fermi energy and diffusion constant can be tuned at will, over a broad carrier density range extending from hole to electron metallic conduction. Theoretical simulations of CF in graphene suggest a possible enhancement of the fluctuation amplitude with respect to standard mesoscopic samples, depending on the strength  or nature of disorder (intervalley scattering) \cite{UCFBeenaker,Altshuler, Kharitonov}. On the experimental side CF have been reported by  several groups \cite{berger,morpurgo,liu,Kechedzhi,folk09,Berezovsky}. But to our knowledge the present work is the first complete investigation of their  correlations and amplitudes as a function of Fermi energy and magnetic field, for both monolayer (ML) and  bilayer (BL) graphene. The importance of the comparison lies in the fact that whereas ML and BL have similar resistivities and thus mean free paths (see Fig. \ref{Fig3}), the (massless and massive) carriers have different velocities because of the different dispersion relations in these two materials. Thus the diffusion constants and therefore correlation energies and fields will have different carrier density dependences, providing a powerful test of the applicability of the theory of mesoscopic fluctuations in those systems.
We find that the variations with carrier density $n$ of the correlation field and energy  are well related to those  of the diffusion coefficient.  We also find, in contrast with  \cite{liu}, that the amplitude of $n$-dependent fluctuations are largest near the charge neutrality point.

We also measure the second order conductance  fluctuations  \cite {defG2} which, unlike linear conductance, are a probe of Coulomb interaction and screening, an important issue in graphene. Second order CF are inherent to systems lacking spatial inversion symmetry (because of random disorder in diffusive systems or geometry  in ballistic systems). They stem from current-induced changes in the carrier density, which in turn, via Coulomb interaction, modify the electrostatic potential landscape, thereby inducing a current-dependent, or second order, change in the conductance of mesoscopic samples. Unlike the first order conductance which 
is even in field, the second order conductance has an odd part in field which was calculated for ballistic \cite{Buttiker} and  diffusive \cite{Spiv} systems. Whereas this mesoscopic rectification was experimentally investigated in ballistic mesoscopic systems (GaAs/GaAlAs quantum dots, Aharonov Bohm rings, carbon nanotubes \cite{others,angers07}), we provide in this letter the first measurement of second order CF in a diffusive system. We find, in qualitative agreement with theoretical predictions, that the odd in field rectification decreases with conductance.

\begin{figure}
    \includegraphics[clip=true,width=9cm]{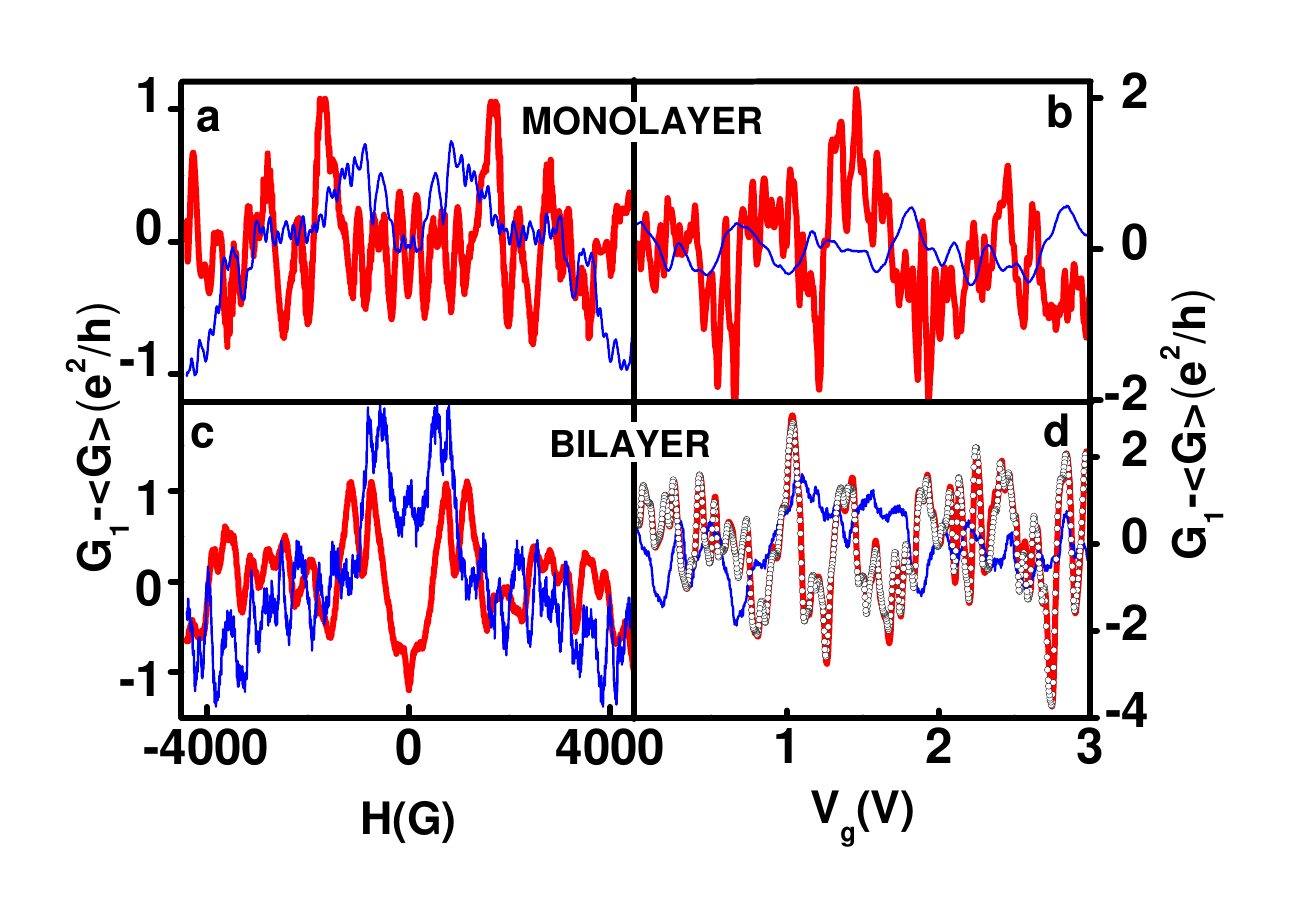}
    \caption{CF as a function of magnetic field (left) and gate voltage (right). A linear fit was subtracted to remove the average conductance. Thick red (resp. thin blue) curves are measured around  $V_g=V_{0}$, the charge neutrality point (CNP) (resp. at high carrier density, corresponding to $V_g - V_{0} = 40.5~V$ for the ML  and $V_g - V_{0} = 15~V $  for the BL). The CNP is at $V_{0} =15.5~V $ for the ML and $V_{0} = -6~V$ for the BL. It is clear (b,d) that the amplitude of $V_g$-dependent CF is larger at the CNP than at high density, both for the ML and the BL. Panel d also shows the reproducibility of the CF, via two nearly indistinguishable successive gate voltage sweeps (open dots and red line). }
    \label{Fig1}
\end{figure}

\begin{figure}
    \includegraphics[clip=true,width=9cm]{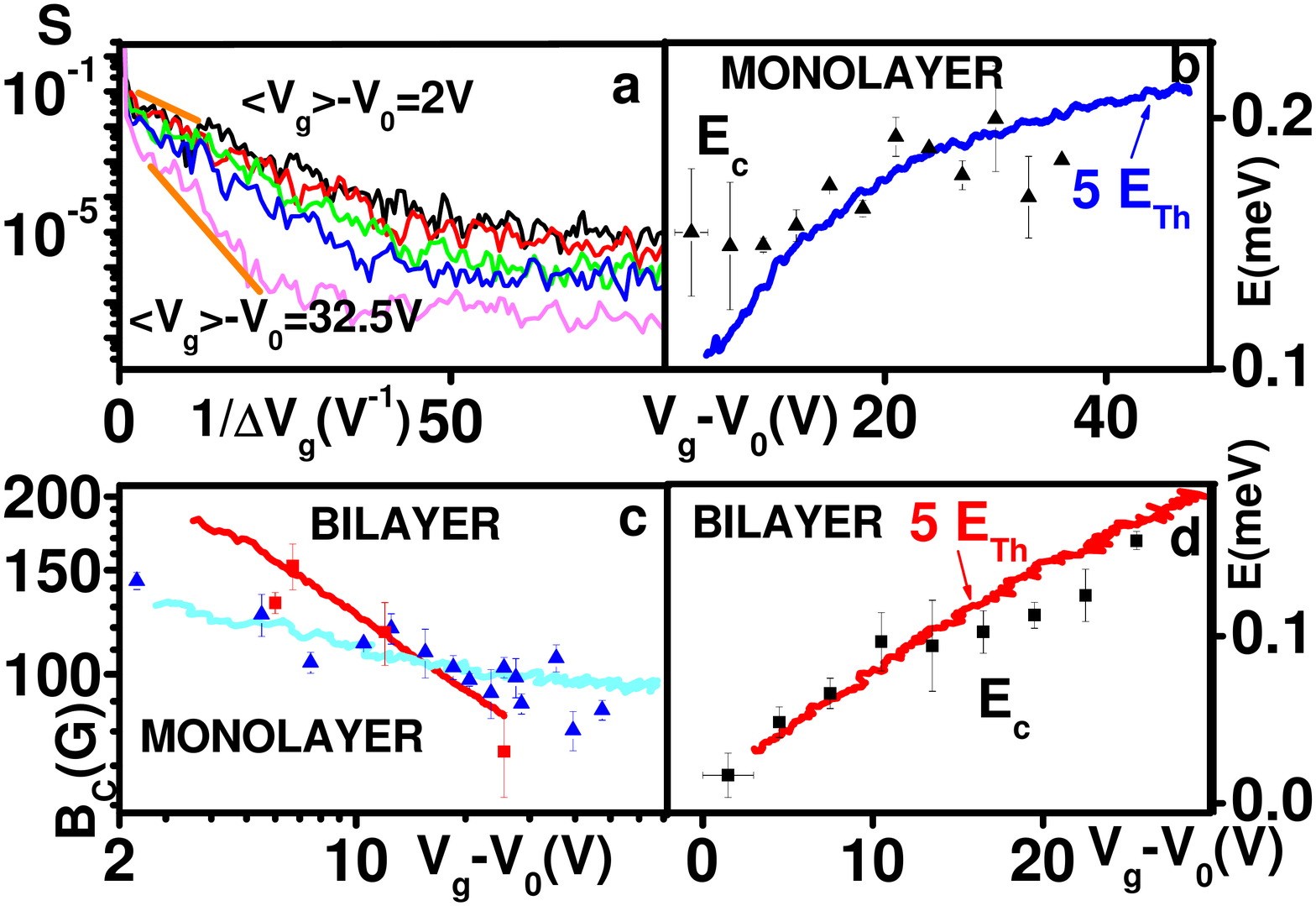}
    \caption{Correlation energies and fields of CF. (a) Fourier power spectrum of CF at different $V_g$ in a semi-log scale. $V_{c}$ is extracted from the exponential fit at small $\Delta V_g^{-1}$. (b) and (d): $E_c(V_g)$ for the ML and BL. Continuous lines are $5E_{Th}$, with $D$ determined from previous magnetoresistance measurements (see text). (c): $B_c(V_g)$. Continuous lines are $B_{\varphi}= 6\Phi_{0}/(L_{T}L)$ for the ML and $4\Phi_{0}/(L_{T}L)$ for the BL, with $D$ similarly determined. 
Vertical error bars represent uncertainty in fit parameters, and horizontal bars the 3 V range used to calculate correlation energies. 
}
    \label{Fig2} 
\end{figure}

\begin{figure}
    \includegraphics[clip=true,width=9cm]{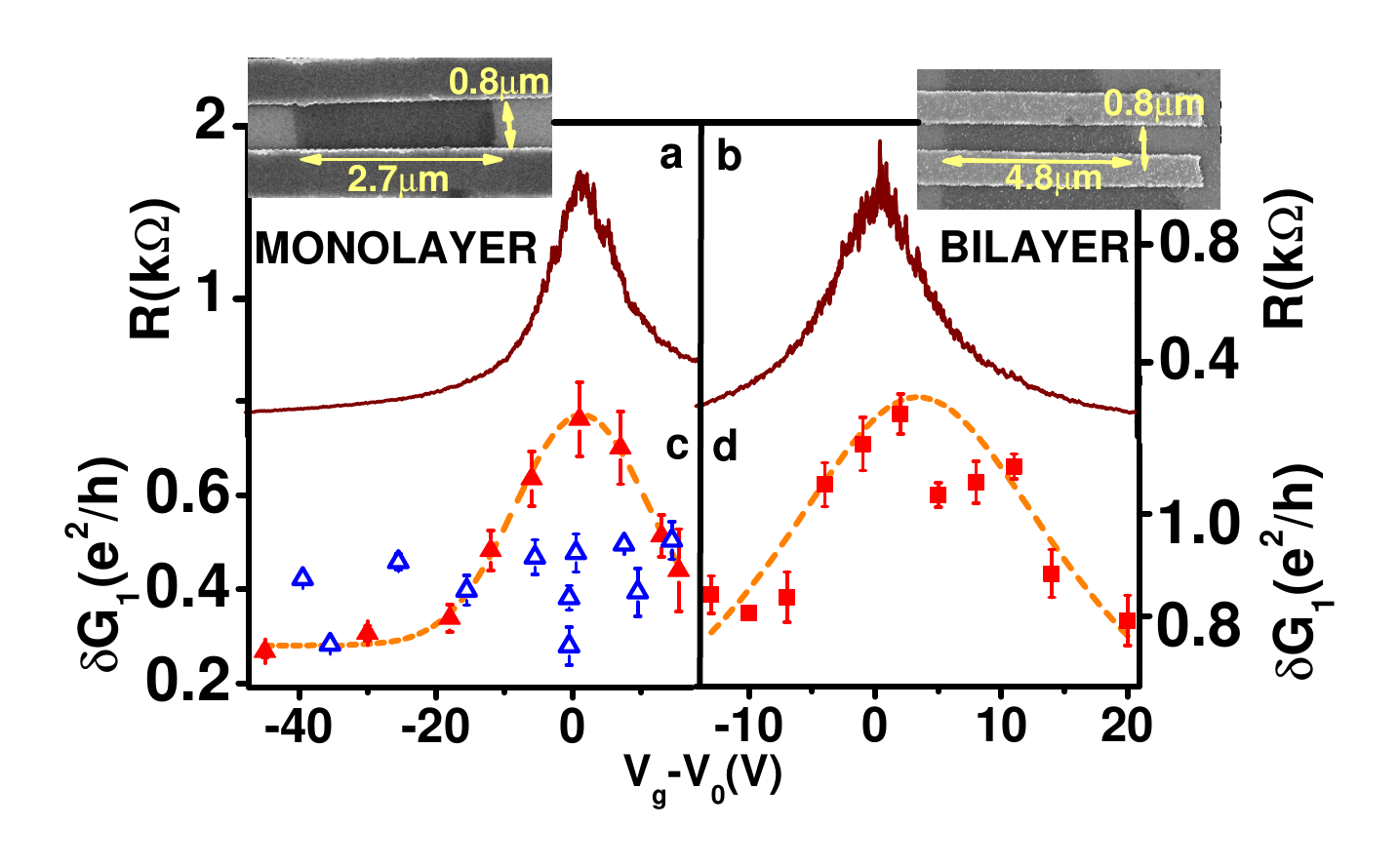}
    \caption{Comparison between resistance and CF. The $V_g$-dependent fluctuations (filled symbols) are larger near the CNP for both the ML and BL (panels c and d). Error bars are the standard deviation of the difference of two different sweeps. Dashed lines are guides for the eyes. The $V_g$-dependent fluctuation amplitude changes with $V_g$ in a qualitatively similar way as the resistance. The $B$-dependent CF amplitude (open triangles) doesn't change much with $V_g$.}
    \label{Fig3}
\end{figure}

Both ML and BL graphene samples (the nature of which was confirmed by Raman spectroscopy) were exfoliated and deposited onto doped silicon substrates with a 285 nm thick oxide. The electrodes  were fabricated by electron beam lithography and sputter deposition of 40 nm thick palladium, providing low  contact resistances ($200~\Omega$ for the ML and $20~ \Omega$ for the BL). The dimensions are $W=2.7~\mu m$, $L=0.8~\mu m$ and $W=4.8~\mu m$, $L=0.7~\mu m$  for the ML and BL, respectively (Fig. \ref{Fig3}). The carrier density $n$ is controlled for both systems by the voltage $V_g$ applied to the doped silicon back gate, via $n=k_F^2/\pi=C(V_g-V_0)$, with $C$ the gate capacitance per unit area, $k_F$ the Fermi wave vector, and $V_0$ the gate voltage of the charge neutrality point. The transport parameters of both samples were determined via classical magnetoresistance measurement, described in \cite{monteverde}. We found that the transport mean free path $l_{tr}$ of both ML and BL varies between 25 and 100 nm over the $60~V$ gate-voltage range probed. Thus $l_{tr}$ is much smaller than the sample dimensions, so that both samples are in the diffusive regime.
Two terminal resistance measurements were performed at 60 mK in a dilution refrigerator with  resistive lines  filtered  at room temperature. The ac current amplitude $I_0$ was adjusted between 5 and 50 nA, depending on the sample resistance, to avoid  heating while optimizing the signal to noise ratio. The first and second harmonics $V_1$ and  $V_2$ of the ac voltage were measured with a low noise voltage preamplifier and a lock-in amplifier. The first and second order conductances were determined via $G_1 = I_0/V_1$ and $G_2 = 2V_2 I_0/V_1^3$ \cite{defG2}. The magnetic field was limited to less than 0.4 T, so that the contribution of Shubnikov de Haas oscillations is negligible. The CF of $G_1$ were measured as a function of B and $V_g$, for both the ML and BL (see Fig. \ref{Fig1}). The  $V_g$ dependent fluctuations were analyzed in 3 V-wide windows, over which the density $n$ can be considered practically constant. The fluctuations of $G_1$, even in $B$ as expected in a two probe measurement, were characterized by their correlation functions and their amplitude $\delta G_1$, defined as the square root of their variance. The histogram of these fluctuations is Gaussian over the whole parameter range.  
The correlation gate voltage and field were determined in the following way. We calculated the Fourier power spectrum of the fluctuations as a function of $k$, the conjugate variable of $B$ or $V_g$, for each set of field or gate voltage data corresponding to a given  average carrier density. This spectrum is the Fourier transform of the correlation function of the fluctuations. Each dataset yielded exponential-like functions which correspond to a lorentzian correlation function in  the direct space (Fig.  \ref{Fig2}a). The decay $exp(- k x_c)$ at low $k$   yields the correlation gate voltage  or field   $x_c$  for each  dataset. The correlation energy $E_c$ is related to  the $V_g$ correlation scale  $V_c$ by  $E_c = V_c \hbar v_F\sqrt{\pi C/(4e V_g)}$ for the ML and $ \hbar^2 \pi CV_c/(2em_{eff})$ for the BL, with $m_{eff}=0.03m_e$ the (constant) low energy effective mass in the BL. These relations are incorrect near the charge neutrality point (CNP) because of density inhomogeneities, so that we deduced $E_c$ from $V_c$ via $k_F(V_g)$ determined in \cite{monteverde}.
   
 In Fig.\ref{Fig2} we compare the variations of $E_c$ and $B_c$ to the predictions \cite{lee87}  $ E_{\varphi}=\hbar D/L_{min}^2$ and $B _\varphi =  \Phi_0 / (L_{min} W_{min})$,  where $L_{min} = min (L, L_T, L_{\varphi})$ and $W_{min} = min (W,L_T,L_{\varphi})$ correspond to the typical longitudinal and transverse length of interfering trajectories; $L_T = \sqrt{\hbar D/k_BT}$ and $L_{\varphi}$ are respectively the thermal  and the phase coherence length.  We first determine the  diffusion coefficient $D=\sigma/(e^2 \nu(E_F))= v_{F}l_{tr}/2$, where $\sigma$ is the conductivity and $\nu(E_F)$ is the density of states,  $v_F$ is constant ($10^6$ m/s) for the ML and is $\hbar k_{F}/m_{eff}$  for the BL. Both $l_{tr}$ and $k_F$  were extracted in \cite{monteverde}. We find that at $60~mK$  $L_T $   varies between  $0.7$ and $1.8~\mu m$, so that $L<L_T <W$  for both samples over the entire density range investigated. Thus $W_{min}= L_T$ and  $L_{min} = L$. Consequently $E_c$ is expected to vary like the Thouless energy $E_{Th}= \hbar/\tau_D= \hbar D/L^2$, and $B_c$ like $\Phi_0/(L L_T)$ \cite{notes}. As shown in Fig. 2, this is indeed what is found: the correlation energy deduced from the experimental data follows $E_c=5~E_{Th}$, and the correlation field follows $B_c=6~\Phi_0/(L L_T)$ for the ML and $B_c=4~\Phi_0/(L L_T)$ for the BL. In particular, we find the expected linear dependence of $E_c$ with $D$, which itself varies like $V_{g}$  for the BL and $\sqrt{V_{g}}$  for the ML. This stems from the fact that in both samples the conductance varies  linearly in $n \propto V_g$ within logarithmic corrections    \cite {monteverde}, and that $\nu(E_F)$ is independent of $V_g$ for the BL and varies like $\sqrt{ V_g}$ for the ML. Similarly  $B_c\propto\Phi_0/L L_T$ is expected to vary like $D^{-1/2}\propto V_g ^{-1/2}$ and $V_g ^{-1/4}$,  as observed respectively for the BL and ML. The numerical factors may be explained by the samples' aspect ratios, but more theory is needed to assess this point.

We now discuss the CF amplitude. As is visible in figures \ref{Fig1}b and \ref{Fig1}d, the $V_g$-dependent CF are greater near the CNP. This is quantified by the fluctuation  amplitude $\delta G_1$ plotted as a function of $V_g$ in Fig. \ref{Fig3}. 
In contrast, the $B$-dependent fluctuations do not significantly depend on the density $n$. This  non ergodicity of the fluctuations in graphene may be a consequence of the  spatial inhomogeneities of $n$ close to the CNP.
Indeed, in a good conductor (large g), changing the Fermi energy is equivalent to changing the disorder configuration, and induces CF of order $e^2/h$. In graphene on the other hand, it has been shown \cite{Yacoby} that near the CNP the system breaks into conducting puddles of electrons and holes, and that
transport takes place along an intricate percolating network of these n and p-type regions \cite{Altshuler}. Thus, as pointed out in \cite{UCFBeenaker}, the change in configuration induced by a change of $V_g$ in this region may induce a much larger variation of conductance than a change in the disorder configuration in a good conductor. In contrast, the magnetic field does not affect the network but only the phases of the wave functions, which explains why  the fluctuations with B are independent of gate voltage \cite{UCFBeenaker}. Note that these results are in contradiction with those of \cite{liu} who found a smaller CF amplitude close to the CNP. This discrepancy could be due to a different nature of disorder in the samples \cite{UCFBeenaker}.
For the ML, we find the $B$-dependent CF amplitude to be $0.4\pm 0.1~e^2/h$, independent of $V_g$. Theory \cite{lee87} predicts $ \delta G = 0.7\sqrt{W/L}~e^2/h$ when  the distance between electrodes L is smaller than $L_{T}$.  This yields  $\delta G = 1.2~e^2/h$ for the  ML, which is three times the value measured in this work (Fig. \ref{Fig3}).

\begin{figure}
    \includegraphics[clip=true,width=7cm]{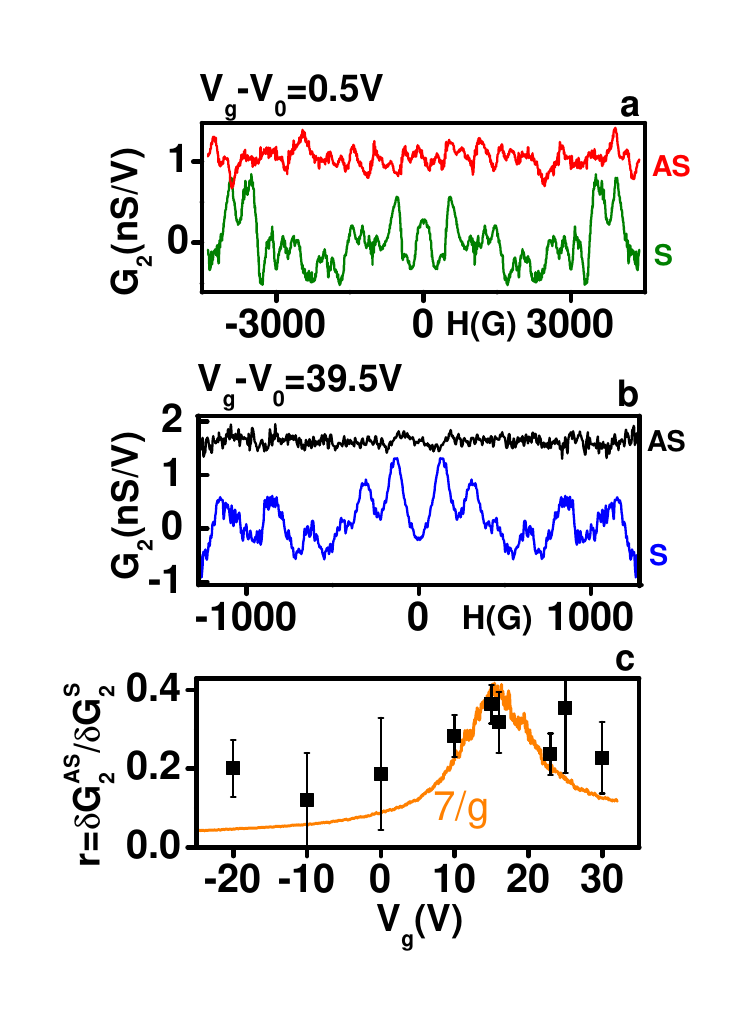}
    \caption{Second order conductance fluctuations for the ML, near (a) and far (b) from the CNP. The antisymmetric part $G_2^{AS}=(G_2(H)-G_2(-H))/2$ is largest at the CNP. (c)The ratio $r=\delta G_{2}^{AS}/\delta G_{2}^{S}$ is compared to $7/g$. Error bars are deduced from the difference of $G_{2}(H)$ between two consecutive scans, which is mainly due to telegraphic noise in the sample.} 
\label{Fig4}
\end{figure}

We now turn  to second-order non-linear CF in monolayer graphene. As mentioned in the introduction, current through a mesoscopic sample induces charge accumulation around impurities and sample edges. This bias  induced change of  electronic density $\delta n(r,V)$  modifies (via Coulomb interactions) the electrostatic  potential landscape throughout the sample by $\delta U_{dis}(\gamma_{int},V)$, where $\gamma_{int}$ quantifies e-e interactions \cite{christen}. As a result, CF acquire a  non linear  bias dependent contribution : $G=G[U_{eq}]+\delta G[\delta U_{dis}(\gamma_{int},V)]$, where $U_{eq}$ is the potential in zero bias. $G_{1}=G[U_{eq}(r)$ is the linear conductance, and $G_{2}=(\partial G[\delta U_{dis},V,\gamma_{int}]/\partial V)_{V=0}$ is the second order conductance.  In a magnetic field $B$, $\delta n(r,V)$ and $\delta U_{dis}(\gamma_{int},V)$ contain  a component which is even, as well as another which is odd in $B$, which can be viewed as a mesoscopic Hall voltage.  Contrary to $G_1$ which is even in $B$, in the presence of interactions  $G_2$ has a component which is odd in $B$, $G_{2}^{AS}$.
The  ratio between the even (symmetric) variance $\delta G_{2}^{S}$ and the odd (antisymmetric) variance $\delta G_{2}^{AS}$ of $G_{2}$  is predicted to be independent of conductance in ballistic systems, but in diffusive systems it should vary like
$\frac{\delta G_{2}^{AS}}{\delta G_{2}^{S}}= \frac{\gamma_{int}}{g}$, with g the conductance in units of $e^2/h$ \cite{ Spiv, polianski}. 
Figure \ref{Fig4} shows $G_{2}^{S}$ and $G_{2}^{AS}$ at two different carrier densities (close and far from the CNP), as well as the ratio of the odd to even amplitude $r=\frac{\delta G_2^{AS}}{\delta G_2^S}$. We find that $r$ significantly decreases as carrier density increases, in contrast to ballistic GaAs/GaAlAs rings \cite{angers07}, where $r$ was  nearly  independent of  the dimensionless conductance $g$. 
 To compare to the predictions for diffusive systems, $r=\gamma_{int}/g$, we plot in Fig. \ref{Fig4}c $r(V_{g})$, and find that $r\approx (7\pm 1)/g $. According to \cite{Buttiker}, the interaction constant $\gamma_{int}$ is related to the geometrical capacitance $C$ and electro chemical capacitance $C_\mu = (1/(\nu e^2) + 1/C)^{-1}$ by $\gamma_{int} = C_\mu/C$. Thus $\gamma_{int} =1/(1+C/(\nu e^2))$ \cite{Buttiker}. Perfect screening (in materials with large densities of states) corresponds to $\gamma_{int}=1$, and no screening to $\gamma_{int}=0$. In graphene $\gamma_{int}\approx 1$ over the entire $n$ range investigated,  within  less than $4/1000$ : screening is strong, even close to the CNP  where $\gamma_{int} = 0.996$. The factor $7$, compared to $1$ expected for $L=W$, may be due to the large aspect ratio $W/L$ of the sample, which is known to increase $\delta G_1$ entering in the calculation of $r$ \cite{angers07}, but there are no calculations for our geometry.
  
In conclusion we have shown that mesoscopic graphene samples exhibit conductance fluctuations which Fermi energy- and $B-$dependent correlation functions can be described by theoretical predictions for diffusive systems over a wide range of carrier concentration,  for both monolayer and bilayer. The different behaviors of the correlation energy and fields are intimately related to the fundamentally different dispersion relations of both systems. A significant increase of the amplitude of the Fermi energy-dependent fluctuations is observed close to the neutrality point, whereas the $B$-dependent fluctuation amplitude is nearly constant over the entire carrier density range. This non-ergodicity of fluctuations may be attributed to the particular disorder due to electron and hole puddles in graphene near the Charge Neutrality Point. Finally, we have measured the second order non-linear conductance. We have exploited the tunability of graphene's conductance to find that its field asymmetry decreases with g, in agreement with theoretical predictions for diffusive systems. This indicates strongly screened electron-electron interactions in graphene.
 
\section{Acknowledgments}
We acknowledge useful discussions with B. Altshuler, A. Chepelianskii, J.Basset, G. Montambaux, M. Polianski, C. Texier, and M. Titov. C.O-A. is
funded by CEE MEST Program CT 2004 514307 EMERGENT CONDMATPHYS Orsay, and RTRA Triangle de la Physique. M. Monteverde was financed by the EU "`Hyswitch" grant  and the CNano Ile de France programm.


\begin{thebibliography}{99}
\bibitem{notes}  We have assumed $L_{\varphi} > L_T$, which is always true when phase coherence is limited by electron-electron interactions.  Note also that ref. \cite{Kechedzhi} explores the different, high temperature regime where $E_\varphi =k_BT$.
\end{thebibliography}
\end{document}